\documentclass[10pt,aps]{revtex4}
\usepackage{amssymb}
\usepackage{latexsym}
\usepackage{epsfig}
\usepackage{float}
\usepackage{graphicx,epsfig,color}
\usepackage{graphicx}
\usepackage{subfigure}
\usepackage{hyperref}
\usepackage{amsmath}

\begin{document}
\title{\textbf{QCD phase transition with non-extensive NJL model in the strong magnetic field}}
\author{Jia Zhang\footnote{202212602002@email.sxu.edu.cn} 
and Xin-Jian Wen\footnote{wenxj@sxu.edu.cn}  }

\affiliation{Institute of Theoretical Physics, State Key Laboratory
of Quantum Optics and Quantum Optics Devices, Shanxi University,
Taiyuan, Shanxi 030006, China}

\begin{abstract}
In this work we make use of the Nambu--Jona-Lasinio model to investigate 
thermodynamic properties of magnetized three-flavor quark matter. 
The non-equilibrium Tsallis distribution is characterized by a dimensionless 
non-extensive parameter $q$. We find that the system always undergoes a 
crossover transition for all given $q$ values and the pseudo-critical temperature 
decreases with the increasing non-extensive parameter. We show that 
the variation of the pressure and its anisotropy in parallel and perpendicular directions. 
However, the diamagnetic nature appears within a certain temperature range 
at zero chemical potential. Moreover, strong magnetic fields can affect the property of 
a non-extensive system by changing the magnetization significantly. 
At high temperatures, the paramagnetic nature always occur no matter how strong 
the magnetic field is. Finally, we study the effects of non-extensive parameter $q$ 
on the trace anomaly and the speed of sound of the system.
\end{abstract}

\maketitle

\section{Introduction}
As a non-Abelian gauge theory, quantum chromodynamics (QCD) contains 
abundant properties with fewer parameters, especially multiple 
phase structures, and is playing a crucial role in the exploration of 
particle physics and astrophysics\cite{Peskin:1995ev,Satz:2012zza,Gross:1973id}. 
The Landau phase transition is the study of the transition between phases with or 
without a certain symmetry, and the specific transition process 
described in terms of order parameter is typically accompanied by 
drastic changes in matter or physical properties. In general, 
order parameter defined as control parameter that allows two phases 
to be distinguished by non-analytic behavior is not unique for a 
system\cite{Borsanyi:2010bp}. As one of the most significant phase transitions, 
the chiral phase transition is related to the mass origin of 98 percent of 
the visible matter in the universe\cite{Fukushima:2010bq}. Furthermore, 
the pseudo-critical temperature of the phase transition is a fundamental 
physical quantity of QCD theory that significantly contributes to 
our comprehension of both the phase diagram and 
QCD theory itself\cite{Borsanyi:2010cj,Borsanyi:2010zi}.

The phase diagram is also influenced by external parameters, such as 
magnetic field, which presents an intriguing area for both experimental and 
theoretical exploration\cite{Mizher:2010zb,Chernodub:2011mc}. Due to 
its profound impact across all regions of the phase diagram, understanding 
QCD matter under strong magnetic fields has become one of the most relevant topics 
in contemporary physics. Indeed, certain compact stars are expected to 
possess strong magnetic fields, and the magnetic fields in their interior are 
much higher than on the surface\cite{Kouveliotou:1998ze}. It is also 
an established fact that there are strong magnetic fields in 
the Relativistic Heavy Ion Collider (RHIC) and the Large Hadron Collider (LHC), 
generating strong interacting quark gluon plasma 
(QGP)\cite{Westfall:1976fu,Warringa:2012bq,Bzdak:2011yy,Fukushima:2010vw}. 
Here, it should be emphasized that magnetic fields do play a crucial role in 
the dynamics of numerous relativistic and quasi-relativistic systems. These range from 
the evolution of the early universe, as already mentioned, to the fireball of 
QGP produced by heavy ion collisions, from compact stars to 
countless systems of condensed matter\cite{Vachaspati:1991nm}. The impacts of 
magnetic field on deconfinement and the chiral transition at finite temperature 
have been explored in lattice 
QCD\cite{Endrodi:2019whh,Meyer:2015wax,Ilgenfritz:2013ara,Bornyakov:2013eya,DElia:2010abb}.

We know that in the middle and low energy region of the phase diagram, 
perturbative QCD is no longer applicable, and the existence of the sign problem 
also makes the lattice QCD unable to perform first-principle calculations 
for relevant non-perturbative problems\cite{Ukawa:2015eka}. Inspired by 
the BCS superconductivity theory in condensed matter physics, in 1961 Yoichiro Nambu and 
G. Jona-Lasinio proposed a now widely used 
non-perturbative model\cite{Nambu:1961tp,Nambu:1961fr}. 
The Nambu--Jona-Lasinio (NJL) model was initially employed to describe 
the known field theory of elementary particles--nucleons, and was later used as a valid 
quark theory model\cite{Buballa:2003qv}. The construction principle of 
the Lagrangian density aims to preserve the symmetry of QCD as much as possible, 
and the most important symmetry is chiral symmetry\cite{Ghosh:2005rf}. In order to 
solve the divergence problem caused by non-renormalization in the NJL model, 
three-dimensional momentum cutoff, four-dimensional momentum cutoff and 
dimension normalization schemes are typically utilized. The deconfinement phase transition 
associated with Z(3) central symmetry breaking has been generated in PNJL model 
at finite temperature, and recently has been extended to zero temperature 
in PNJL0 model\cite{Fuseau:2019zld,Ratti:2005jh,Wang:2022xxp}.

As mentioned above, heavy ion collisions occurring in 
RHIC and LHC evolve so quickly that the entire system is far from 
equilibrium state\cite{Randrup:2009gp}. Therefore, we cannot use 
the usual Boltzmann-Gibbs (BG) statistic\cite{Megias:2015fra} to study phase space. 
In 1988, Constantino Tsallis first proposed the Tsallis statistic as a generalization of 
the BG statistic and gained widespread attention\cite{Tsallis:1987eu}. He investigated 
a non-extensive form of entropy and constructed systems characterized by 
the real parameter $q$\cite{Tsallis:1998ws}. In such a way that when $q\rightarrow1$, 
the extensive nature of the entropy is recovered. In recent years, many studies 
have taken the Tsallis statistic into account to describe high energy 
physics\cite{Megias:2015fra,Cardoso:2017pbu,Deppman:2016fxs,Lavagno:2011zm,Wilk:2008ue}. 
It is worth noting that for the choice of $q$-value, in principle we can choose 
any value greater than 1, but in practice calculations we choose the typical values 
in the range of 1--1.2 obtained by fitting the transverse momentum distribution in high-energy 
collisions\cite{Azmi:2015xqa,Cleymans:2012ya,Beck:2000nz,Azmi:2014dwa,Bediaga:1999hv}.

The paper is organized as follows. In Section \ref{sec:model},
we introduce Tsallis distribution and show detailed formalism 
of the NJL model in the background of strong magnetic fields.
In Section \ref{sec:result}, we obtain the phase transition in non-extensive state 
and discuss related thermodynamic properties of quark matter. 
Finally we present our conclusions in the last section.

\section{ non-extensive NJL model in the strong magnetic field}\label{sec:model}
The starting point of the NJL model as an effective theory is 
the Lagrangian density to model dynamical chiral symmetry breaking.
In the presence of strong external magnetic fields, the Lagrangian
density of the three-flavor NJL model can be expressed as\cite{Menezes:2008qt}
\begin{equation}
{\mathcal{L}}_{NJL}=\bar{\psi}(\textrm{i}/\kern-0.5emD-m)\psi
+G[(\bar{\psi}\psi )^{2}+(\bar{\psi}i\gamma _{5}\vec{\tau}\psi)^{2}]
-K\left\{det_{f}[\bar{\psi}(1+\gamma _{5})\psi]+det_{f}[\bar{\psi}(1-\gamma _{5})\psi]\right\} ,
\end{equation}
where $\psi$ represents the quark fields which are fields in Dirac,
colour and flavor space, $m$ stands for the bare fermion mass. 
An essential element $/\kern-0.5emD=\gamma^{\mu}D_{\mu }$ 
and the covariant derivative $D_{\mu }=\partial _{\mu}+\textrm{i}QA_{\mu }$ 
represents the coupling of quarks to electromagnetic fields. 
The $Q$ is the electric charge matrix acting in the flavor space 
$Q=\mathrm{diag}_f(q_u,q_d,q_s)=\mathrm{diag}_f(2e/3,-e/3,-e/3)$ with $e>0$. 
The coupling constants G and K, and the cut-off $\Lambda$ are determined
by fitting the pion mass and decay constant. 
A sum over flavor and colour degrees of freedom is implicit. 

We next consider the $q$-exponential function 
defined as\cite{Tsallis:1987eu,Tsallis:1998ws,Cleymans:2011in,Azmi:2015xqa} 
\begin{eqnarray}
\exp_q(x)\equiv\left\{\begin{aligned}
(1+(q-1)x)^{1/(q-1)} \quad x>0\\
(1+(1-q)x)^{1/(1-q)} \quad x\leq 0\\\end{aligned}\right. ,
\end{eqnarray}
where $x=\frac{\omega_i-\mu}{T}$ is defined and in the limit $q\rightarrow 1$, 
the standard exponential $\lim\limits_{q\rightarrow 1} 
\exp_q(x) \rightarrow \exp(x)$ is recovered. The above formula is 
the inverse function of the $q$-logarithm given by
\begin{eqnarray}
\ln_q(x)\equiv\left\{\begin{aligned}
\frac{x^{q-1}-1}{q-1} \quad x>0\\
\frac{x^{1-q}-1}{1-q} \quad x\leq 0\\\end{aligned}\right. .
\end{eqnarray}

Correspondingly, the distribution function of fermions 
at the temperature $T$ and chemical potential $\mu$ is 
\begin{eqnarray}
f_+(\omega_i)=\left\{\begin{aligned}
\frac{1}{1+(1+\frac{(q-1)(\omega_i-\mu)}{T})^{\frac{1}{q-1}}} \quad x>0\\
\frac{1}{1+(1+\frac{(1-q)(\omega_i-\mu)}{T})^{\frac{1}{1-q}}} \quad x\leq 0
\\\end{aligned}\right. ,
\end{eqnarray}
where $\omega_i=\sqrt{p_z^{2}+M^{2}+2k_{i}|q_{i}|B}$ 
gives the effective energy of single particle 
in strong magnetic fields\cite{Bali:2014kia,Endrodi:2013cs}.

In the mean-field approximation(MFA)\cite{Buballa:1998pr,Carroll:2009gwk}, 
the pairing of quark-antiquark with the same chirality results in 
the formation of condensation that can be regarded as the order parameter, 
and the effective dynamical mass can be obtained as
\begin{equation}
M_i=m_i-4G\phi_{i}+2K\phi_{j}\phi_{k}  ,
\label{eq:gap}
\end{equation}
where the quark condensates include $u$, $d$ and $s$ quark contributions
as $\phi=\sum_{i=u,d,s}\phi_{i}$. The effective dynamical mass depends on 
their flavor condensates. 
The contributions from the quark condensates with flavor $i$
include the vacuum, the magnetic field, and the
medium term as\cite{Bali:2012zg,Avancini:2011zz}
\begin{equation}
\phi _{i}=\phi _{i}^{\mathrm{vac}}+\phi _{i}^{\mathrm{mag}}+\phi _{i}^{%
	\mathrm{med}},  \label{eq:condensate}
\end{equation}
where each term reads
\begin{eqnarray}
\phi_i ^{\mathrm{vac}}&=&-\frac{MN_{c}}{2\pi ^{2}}\left[\Lambda \sqrt{%
	\Lambda ^{2}+M^{2}}-M^{2}\ln (\frac{\Lambda +\sqrt{\Lambda ^{2}+M^{2}}}{M})%
\right], \\
\phi _{i}^{\mathrm{mag}} &=&-\frac{M|q_{i}|BN_{c}}{2\pi ^{2}}\left\{
\ln [\Gamma (x_{i})]-\frac{1}{2}\ln (2\pi)+x_{i}
-\frac{1}{2}(2x_{i}-1)\ln
(x_{i})\right\} , \\
\phi _{i}^{\mathrm{med}} &=&\sum_{k_{i}=0}^\infty a_{k_{i}}\frac{M|q_{i}|BN_{c}}{%
	4\pi ^{2}}\int \frac{f_+(\omega_i)+f_-(\omega_i)}{\omega_{i}} dp.
\end{eqnarray}

In order to obtain a good approximation, we choose the dimensionless quantity
$x_i$ is defined as $x_i=M^{2}/(2|q_{i}|B)$ in the renormalized expression.
In Eq.(9), $a_{k_{i}}=2-\delta_{k_i0}$ and $k_{i}$ are the degeneracy label 
and the Landau quantum number, respectively. In nonzero magnetic fields, 
the concepts of Landau quantization and magnetic catalysis, 
where the magnetic field is assumed to be oriented along $z$ direction, 
should be implemented. So we can use the mapping $\int \frac{d^3p}{(2\pi)^3}\rightarrow
\frac{|q_{i}|B}{2\pi}\sum\limits_{k}\int \frac{dp_{z}}{2\pi}(2-\delta_{k_i0})$ 
to compute integral over the three momenta\cite{Fraga:2008qn,Chakrabarty:1996te}. 
It should be noted that the NJL model, as a non-renormalization model, 
needs to be regularization to remove the divergence of the vacuum. In literature, 
there are many works that have discussed the renormalization of the free energy of 
charged fermions in a magnetic field. The appropriate scheme was proposed to 
regularize the integral. Schwinger’s proper time formalism regularized all integrals 
without separating the divergent (vacuum) piece from the convergent (thermomagnetic) contribution. 
The magnetic field independent regularization (MFIR) makes a full separation of 
the finite magnetic contributions and the divergencies, which is employed 
in the present calculation. 

The total thermodynamic potential density 
in the mean-field approximation reads
\begin{equation}
\Omega =\sum_{i=u,d,s}\Omega _{i}+2G\sum_{i=u,d,s}\phi_{i}^{2}-4K\phi_{u}\phi_{d}\phi_{s},
\label{omega}
\end{equation}%
where $\Omega_i$ is defined as $\Omega _{i}=\Omega
_{i}^{\mathrm{vac}}+\Omega _{i}^{\mathrm{mag}}+\Omega
_{i}^{\mathrm{med}}$ in the first term. The vacuum, the magnetic field, and the medium
contributions at finite temperature and chemical potential 
to the thermodynamic potential are\cite{Menezes:2008qt,Rozynek:2015zca}
\begin{widetext}
\begin{eqnarray}
\Omega_i ^{\mathrm{vac}}&=&\frac{N_{c}}{8\pi ^{2}}\left[ 
M^{4}\ln(\frac{\Lambda +\epsilon _{\Lambda }}{M})-
\epsilon _{\Lambda}\Lambda (\Lambda ^{2}+\epsilon _{\Lambda }^{2})\right] ,\\
\Omega _{i}^{\mathrm{mag}}&=&-\frac{N_{c}(|q_{i}|B)^{2}}{2\pi^{2}}\left[
\zeta^{\prime}(-1,x_i)-\frac{1}{2}(x_{i}^{2}-x_{i})\ln (x_{i})+
\frac{x_{i}^{2}}{4}\right],\\
 \Omega _{i}^{\mathrm{med}}&=&-\sum_{k_{i}=0}^\infty a_{k_{i}}
 \frac{|q_{i}|BN_{c}T}{4\pi^{2}}\int dp\Big[
 \ln_q(1+e^{-(\omega_i-\mu)/T})+\ln_q(1+e^{-(\omega_i+\mu)/T})\Big]\notag\\
 &=&\left\{\begin{aligned}
-\sum_{k_{i}=0}^\infty a_{k_{i}}\frac{|q_{i}|BN_{c}T}{4\pi^{2}}\int dp
\left\{\frac{[1+(1-\varphi^{+})^{\frac{1}{1-q}}]^{q-1}-1}{q-1}+
\frac{[1+(1-\varphi^{-})^{\frac{1}{1-q}}]^{q-1}-1}{q-1}\right\} \quad x>0\\
-\sum_{k_{i}=0}^\infty a_{k_{i}}\frac{|q_{i}|BN_{c}T}{4\pi^{2}}\int dp
\left\{\frac{[1+(1+\varphi^{+})^{\frac{1}{q-1}}]^{q-1}-1}{q-1}+
\frac{[1+(1+\varphi^{-})^{\frac{1}{q-1}}]^{q-1}-1}{q-1}\right\} \quad x\leq 0
\\\end{aligned}\right. ,
\end{eqnarray}
\end{widetext}
where the quantity $\epsilon _{\Lambda }$ is defined as $\epsilon
_{\Lambda}=\sqrt{\Lambda ^{2}+M^{2}}$. The $\zeta
(a,x)=\sum_{n=0}^{\infty }\frac{1}{(a+n)^{x}}$ is the
Riemann-Hurwitz zeta function. The
$\zeta'(-1,x)=\frac{d\zeta(z,x)}{dz}|_{z=-1}$ is defined in the
magnetic field term. The function $\varphi^{\pm}$ in the medium term 
is expressed as $\varphi^{\pm}=\frac{(q-1)(\omega_i\mp\mu)}{-T}$ 
as a matter of convenience.

Generally, the pressure can be obtained by $P=-\Omega$. In this context, 
we take care of the normalized pressure by subtracting vacuum pressure contribution 
from the total pressure given as 
\begin{eqnarray}
P_{\mathrm{eff}}=-\Omega-P_0.
\end{eqnarray}
Now, the magnetization of the system is given by
\begin{eqnarray}
{\cal{M}}=\frac{\partial P_{\mathrm{eff}}}{\partial B}.
\end{eqnarray}
where the magnetic field and the medium term are
\begin{eqnarray}
\cal{M}^{\mathrm{mag}}&=&\frac{2P^{\mathrm{mag}}}{B}-
\frac{|q_{i}|N_{c}M^{2}}{4\pi^{2}}\left\{\ln[\Gamma(x_{i})]-
\frac{1}{2}\ln(2\pi)+x_{i}-(x_{i}-\frac{1}{2})\ln(x_{i})\right\} ,\\
\cal{M}^{\mathrm{med}}&=&\left\{\begin{aligned}
\frac{P^{\mathrm{med}}}{B}-\sum_{k_{i}=0}^\infty a_{k_{i}}\frac{|q_{i}|^2BN_{c}k}{4\pi^{2}}
\int dp\frac{1}{\omega_i}\left\{\big[1+\big(1-\varphi^{+}\big)^{\frac{1}{1-q}}\big]^{q-2}
\big(1-\varphi^{+}\big)^{\frac{q}{1-q}}\right. \\
\left.+\big[1+\big(1-\varphi^{-}\big)^{\frac{1}{1-q}}\big]^{q-2}
\big(1-\varphi^{-}\big)^{\frac{q}{1-q}}\right\}   \quad x>0\\
\frac{P^{\mathrm{med}}}{B}-\sum_{k_{i}=0}^\infty a_{k_{i}}\frac{|q_{i}|^2BN_{c}k}{4\pi^{2}}
\int dp\frac{1}{\omega_i}\left\{\big[1+\big(1+\varphi^{+}\big)^{\frac{1}{q-1}}\big]^{q-2}
\big(1+\varphi^{+}\big)^{\frac{2-q}{q-1}}\right. \\
\left.+\big[1+\big(1+\varphi^{-}\big)^{\frac{1}{q-1}}\big]^{q-2}
\big(1+\varphi^{-}\big)^{\frac{2-q}{q-1}}\right\} \quad x\leq 0
\\\end{aligned}\right. .
\end{eqnarray}

The strong magnetic fields along the $z$-axis breaks the Lorentz invariance 
so that the pressure of the system is anisotropic. The contributions 
in the parallel and perpendicular directions can be written in terms of 
the magnetization as\cite{Bali:2013esa}
\begin{eqnarray}
P_{\parallel}&=&-\Omega-\frac{B^2}{2},\\
P_{\perp}&=&-\Omega-{\cal{M}}B+\frac{B^2}{2}.
\end{eqnarray}

Next, from the energy density and the pressure one can easily evaluate 
the trace anomaly (interaction measure), defined as
\begin{eqnarray}
I=\epsilon-3P_{z}+2{\cal{M}}B,
\end{eqnarray}
and the speed of sound squared is essential fundamental quantity that is used 
to describe hot QCD medium, we have
\begin{eqnarray}
c_{s}^{2}=\frac{\partial P}{\partial \epsilon}\equiv \frac{dP/dT}{d\epsilon/dT}.
\end{eqnarray}

\section{Numerical result and conclusion}\label{sec:result}
In the framework of the SU(3) NJL model, 
the following parameters are adopted:  $m_u=m_d=5.5$ MeV and $m_s=135.7$ MeV for 
the current masses of up, down and strange quarks, $\Lambda=631.4$ MeV and 
$G=1.835/\Lambda^{2}$, $K=9.29/\Lambda^{5}$ for the momentum cutoff and 
the coupling constant\cite{Hatsuda:1994pi}. As mentioned earlier, 
the maximum deviation of $q$ should not exceed 20\%, and following this, 
we chose five different $q$ values in the following Tsallis statistic.

\begin{figure}[H]
\centering
\includegraphics[width=0.50\textwidth]{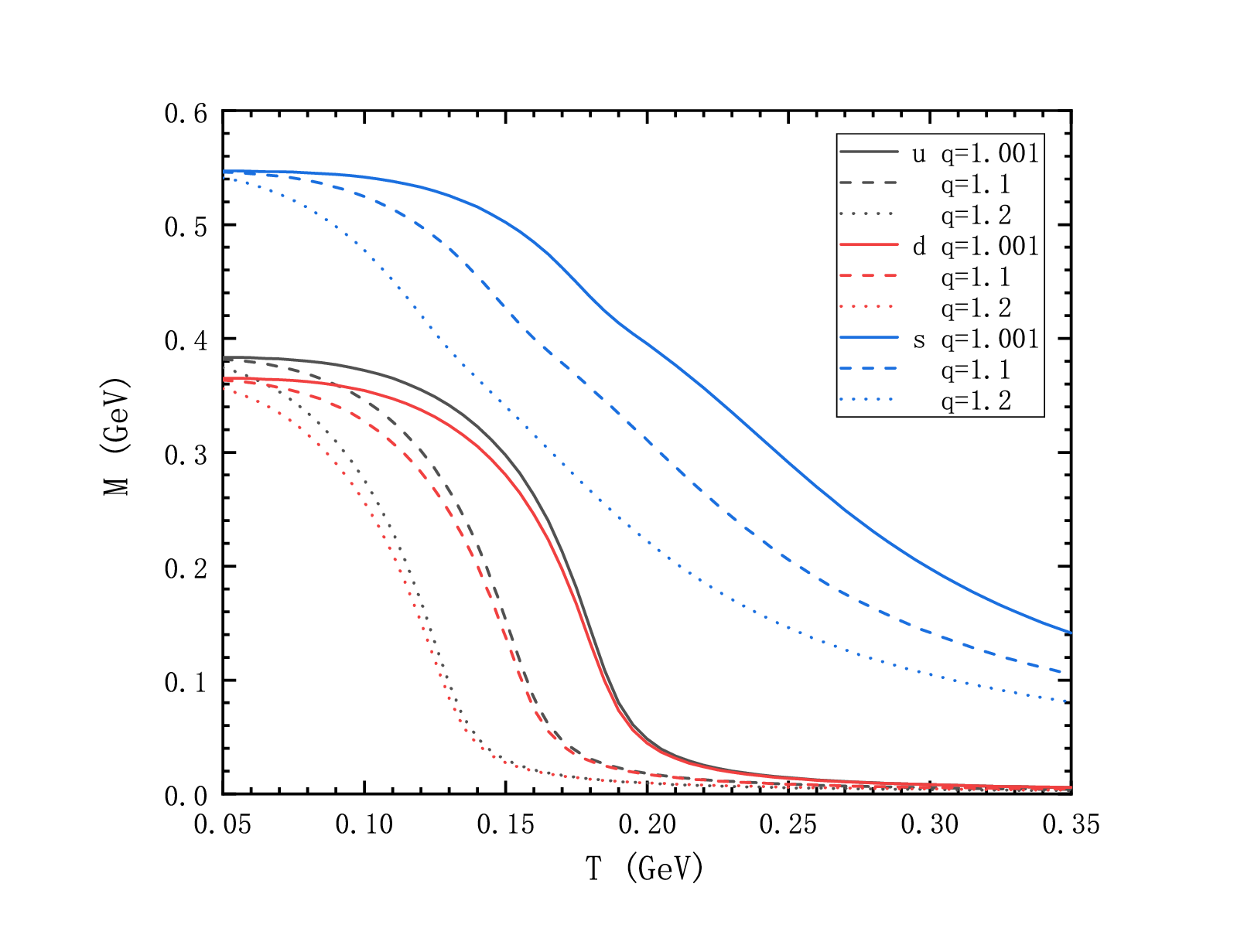}
\caption{The quark dynamical mass as a function of temperature 
for three sets of $q$-value at zero chemical potential and 
the magnetic field $eB=0.2$ GeV$^2$.}\label{fig:Graph1}
\end{figure}

The lattice QCD indicates that the system has no genuine phase transition 
but a crossover at zero chemical potential\cite{Bali:2011qj,Kharzeev:2013jha}. 
In the NJL model, the quark dynamical mass plays as an order parameter to 
signal the chiral crossover. Fig. \ref{fig:Graph1} depicts the dynamical mass of 
$u$, $d$ and $s$ quark in the chiral phase transition at the magnetic field $eB=0.2$ GeV$^2$.
It is clear that the difference of $q$-value does not produce a 
significant change in the quark mass at $T=0$ for each flavor. So it is concluded that
the vacuum mass is almost independent on the parameter $q$.
The curve falls to lower value at high temperature for all given $q$ values, 
which means that quarks are in a chiral restoration. Compared with the $q$=1.2, 
parameter $q$=1.001 leads to a chiral restoration at higher temperature. 
But the decrease of the mass would be more rapidly 
for a larger $q$-value, which indicates that the larger $q$-value pushes 
the chiral restoration to a lower temperature.

\begin{figure}[H]
\centering
\includegraphics[width=0.50\textwidth]{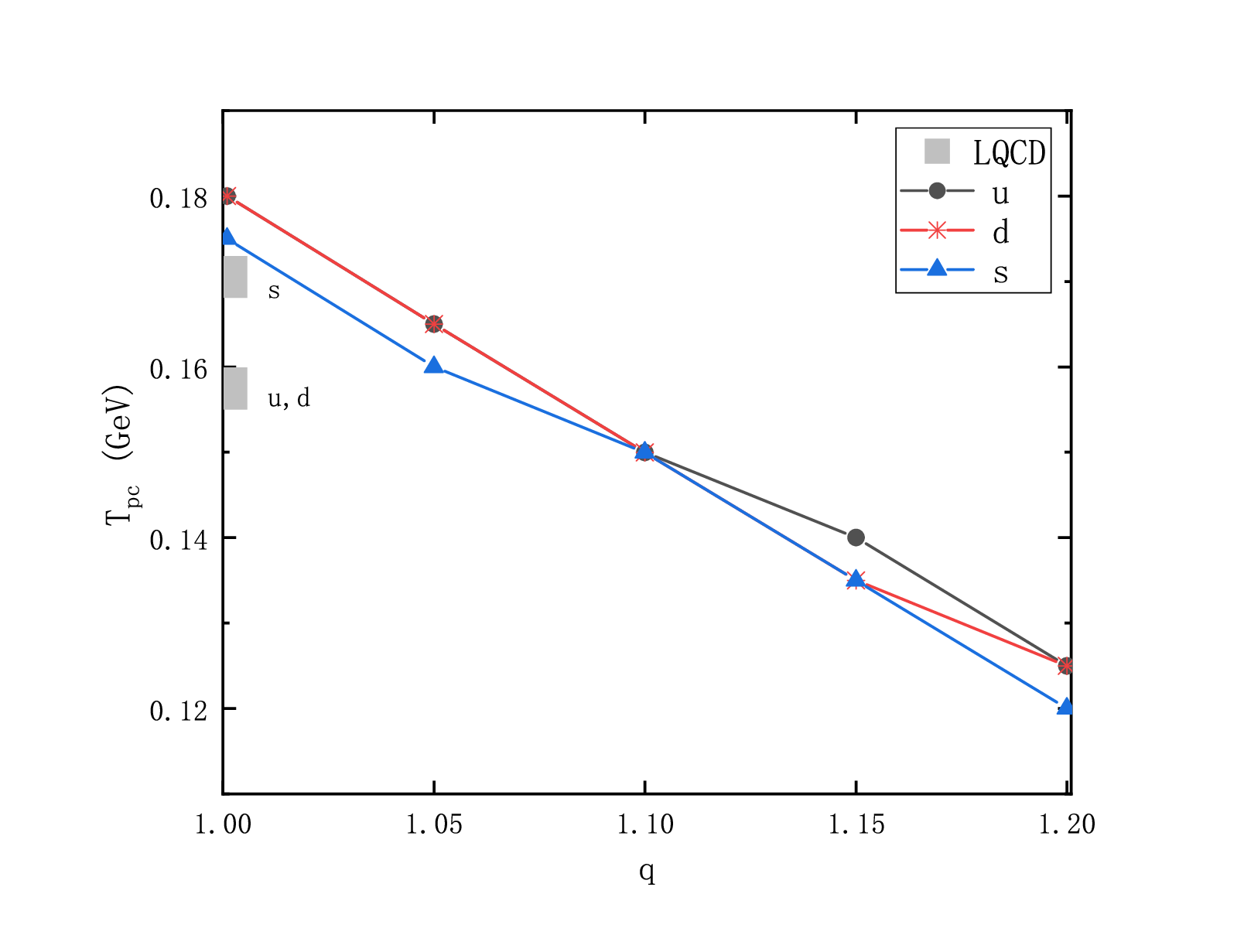}
\caption{The pseudo-critical temperature for the chiral crossover 
as a function of non-extensive parameter $q$.}\label{fig:Graph2}
\end{figure}

Based on the standard method, the pseudo-critical temperature ($T_{pc}$) is obtained 
by the peak of the derivative of the dynamical mass with respect to the temperature. 
We have presented precise data of the pseudo-critical temperature for 
five different non-extensive parameters extracting the values from 
the susceptibility $dM/dT$ in Fig. \ref{fig:Graph2}. The gray area represents 
the equilibrium lattice data in the continuum extrapolation limit\cite{Bali:2011qj}. 
As the value of $q$ increases up to 1.2, the system is far away from equilibrium state 
and the $T_{pc}$ would have a reduction of about 30 percent. In other words, 
non-equilibrium state contributes to the suppression of $T_{pc}$,  i.e, 
the restoration of the chiral symmetry breaking takes place 
at lower temperatures\cite{Shen:2017etj,Zhao:2023xpj}.

\begin{figure}[H]
\centering
\includegraphics[width=0.50\textwidth]{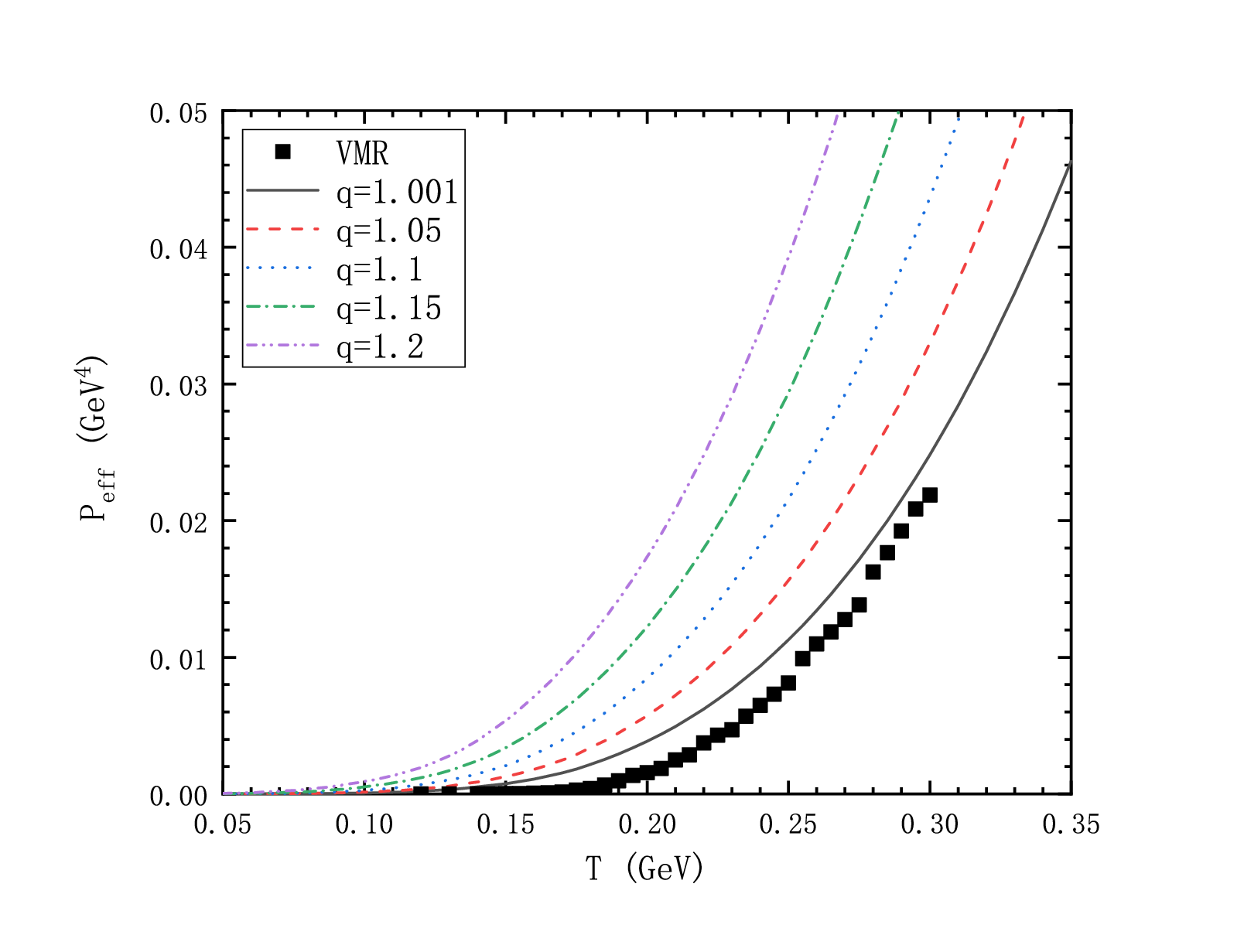}
\caption{The variation of the effective pressure as a function of temperature
for different values of $q$ at $eB=0.2$ GeV$^2$.}\label{fig:Graph3}
\end{figure}

Various thermodynamic quantities can be derived from 
the total thermodynamic potential density in Eq.(\ref{omega}). 
In Fig. \ref{fig:Graph3}, the values of effective pressure were calculated 
for different values of the non-extensive parameter $q$ at vanishing $\mu$. 
We find that pressure monotonically increases as the temperature rises 
for all $q$-values at the magnetic field $eB=0.2$ GeV$^2$. In addition, 
There is a minimum pressure in near equilibrium which is indicated by 
the solid black line of $q$=1.001. An important thing to note is that 
the zero pressure at zero temperature is due to the subtraction of 
the vacuum pressure in the calculation.

\begin{figure}[H]
\centering
\includegraphics[width=0.9\linewidth]{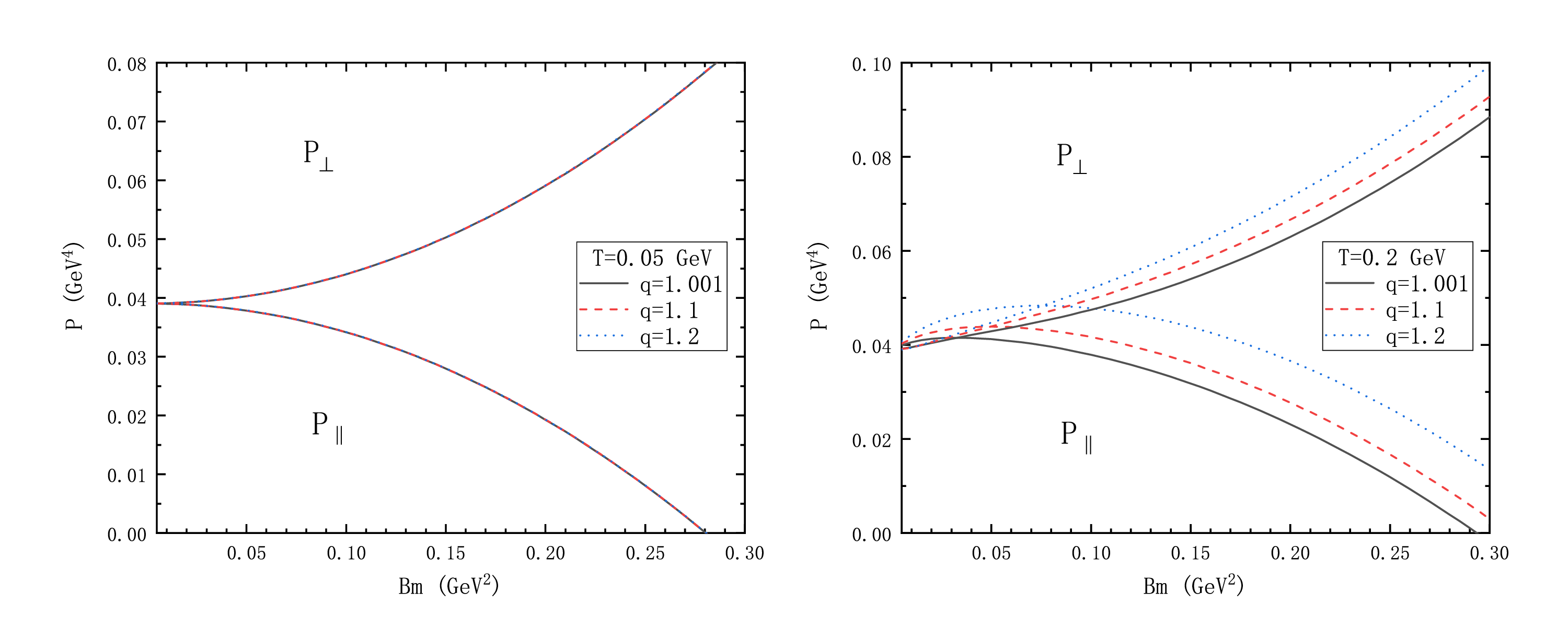}
\caption{The behavior of the transverse pressure and the longitudinal pressure 
for three sets of $q$-value as a function of the magnetic fields 
at $T=0.05$ GeV and $T=0.2$ GeV.}\label{fig:Graph4}
\end{figure}

To investigate the anisotropy of the system in non-extensive state, 
we can change the temperature and observe the variation of the transverse pressure
and the longitudinal pressure for three different non-extensive parameters. 
The behaviors of pressure in the parallel and perpendicular directions are depicted 
as a function of the magnetic fields at $T=0.05$ GeV (left) for the chiral symmetry breaking 
and $T=0.2$ GeV (right) for the chiral symmetry restoration in Fig. \ref{fig:Graph4}. 
One can observe that the longitudinal pressure decreases and the transverse pressure increases 
as the magnetic field strength increases at finite temperature. In the left panel, 
the three $q$ values cases are almost indistinguishable. It can be concluded that 
the effect of the non-extensive parameter is more pronounced in the high temperature regions 
like the right panel. Furthermore, with the rise of the $q$-value, the pressure increases 
in both the parallel and perpendicular directions.

The magnetization is an important thermodynamic quantity to understand QCD matter, 
which reflects its response to an external magnetic field. The sign of 
the magnetization manifests an important magnetic property of 
the system\cite{Ferrer:2019xlr,Cao:2023bmk}. 
The QCD matter with $\cal{M}$\textgreater0 is paramagnetic, where 
the thermal QCD medium aligns towards the direction of the magnetic field. 
On the contrary, the QCD matter with $\cal{M}$\textless0 is diamagnetic and 
the thermal QCD medium aligns oppositely to the direction of the magnetic field and 
produces an induced electric current. In lattice calculation\cite{Bali:2013esa}, 
a good method is provided to calculate magnetization in the presence of magnetic fields
by isolating the magnetic-field-dependent divergent term. The value of magnetization 
could depend on the regularization scheme, which deserves careful study 
for the non-renormalization theory.

\begin{figure}[H]
\centering
\includegraphics[width=0.50\textwidth]{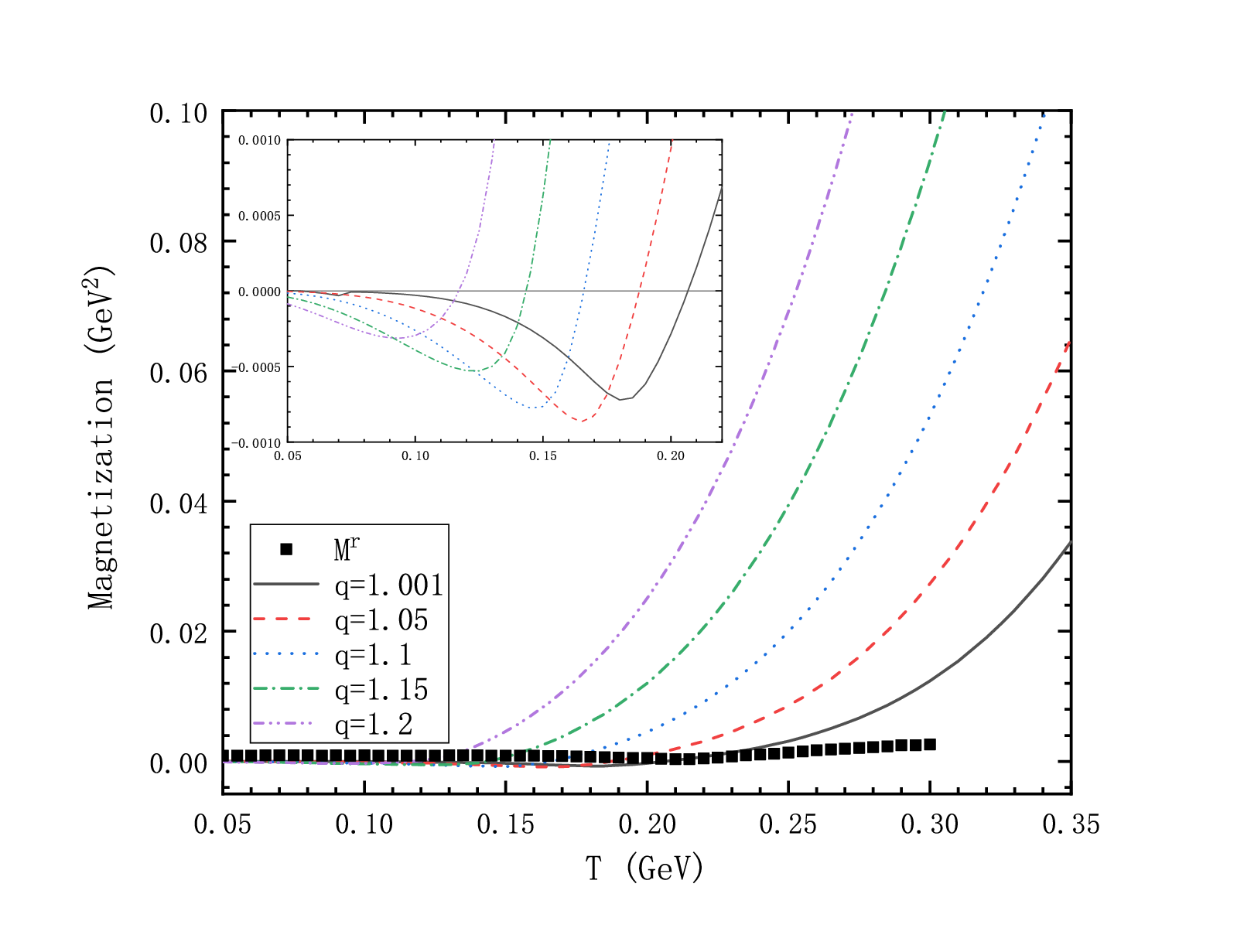}
\caption{The dependence of magnetization on temperature 
for different values of $q$ at $eB=0.2$ GeV$^2$.}\label{fig:Graph5}
\end{figure}

Fig. \ref{fig:Graph5} presents 
the magnetization $\cal{M}$ of the quark matter as a function of temperature 
at vanishing $\mu$ and $eB=0.2$ GeV$^2$. At a cursory glance, the sign of 
the magnetization is positive and the resulting $\cal{M}$ seems to monotonically 
increases with increasing the temperature for all $q$-value. As introduced, 
the positive $\cal{M}$ indicates that the paramagnetic characteristic of 
the quark matter predominates as the temperature increases. The increase of $q$-value 
away from equilibrium would enhance the magnetization substantially within 
a given temperature range. However, inside box of the picture illustrates 
the regional variation of magnetization in relation to temperature for different values. 
It is obvious that the curve is non-monotonic and $\cal{M}$ is negative 
within a specific range. This gives a signature that the system has diamagnetic properties. 
As the temperature is raised, the system reverts to paramagnetic behavior. 
At the same time, there is a minimum value of magnetization for all $q$ values. 
It is interesting that the magnetization in a near equilibrium state shows 
stronger diamagnetic performance compared to the other state far from equilibrium.

\begin{figure}[H]
\centering
\includegraphics[width=0.50\textwidth]{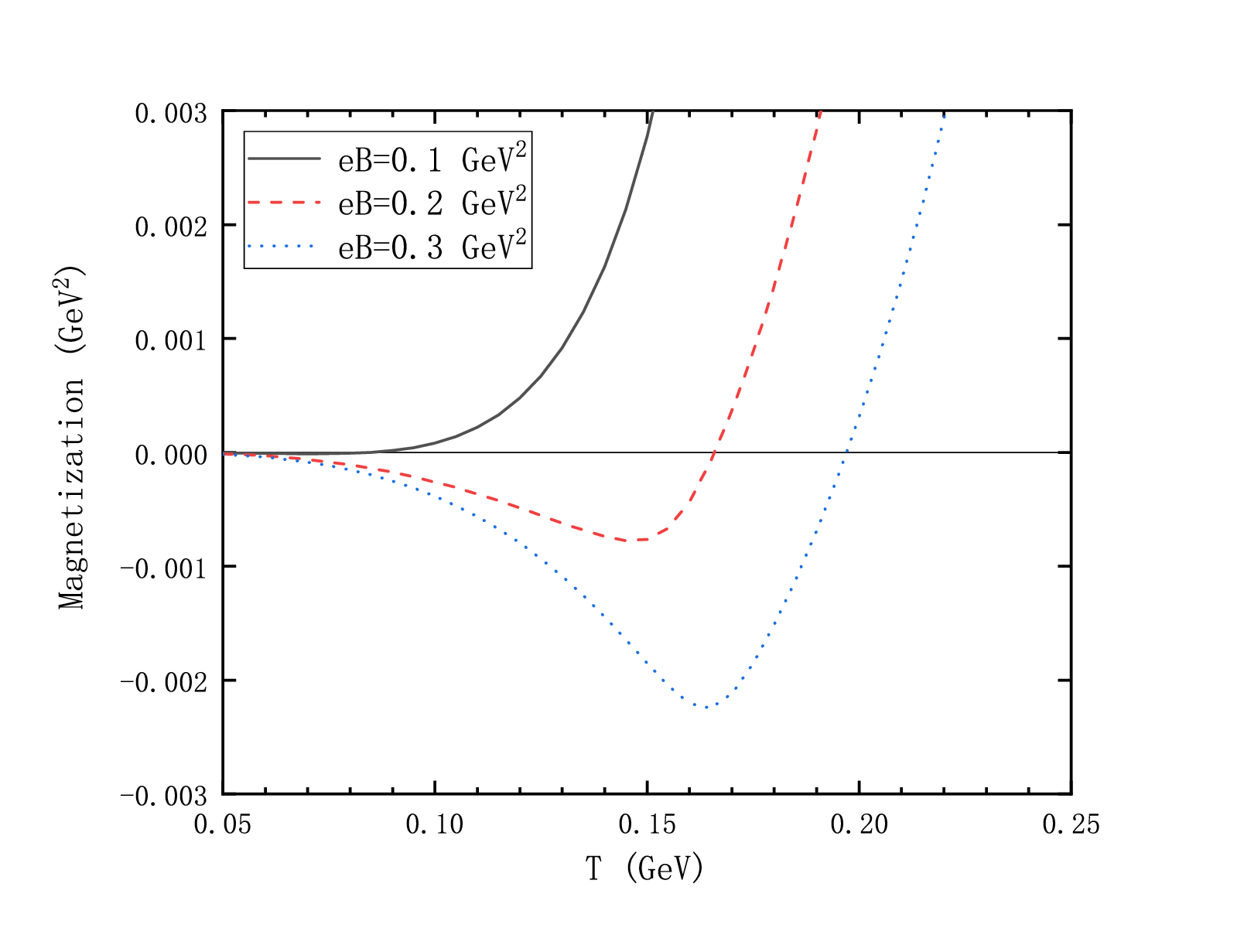}
\caption{The dependence of magnetization on temperature 
for different magnetic fields at vanishing $\mu$ at $q$=1.1.}\label{fig:Graph6}
\end{figure}

In order to show the dependence of magnetization on the magnitude of 
magnetic fields, the magnetization is evaluated as a function of the temperature 
at different magnetic fields in Fig. \ref{fig:Graph6}. We set $q$=1.1 as a 
specific value representing the non-equilibrium state and three magnetic fields 
$eB=0.1,0.2,0.3$ GeV$^2$. The black solid line reflects that the magnetization 
increases monotonically with temperature at small magnetic field $eB=0.1$ GeV$^2$. 
The rise of the magnetic field leads to the occurrence of the non-monotonic behavior 
of the magnetization. At the magnetic field $eB=0.3$ GeV$^2$ marked by 
the blue dotted line, the negative magnetization exists in a larger range of temperature. 
There is a local minimum value of magnetization appearing at a proper temperature 
for larger magnetic fields. Moreover, the minimum value decreases with the increase of 
the magnetic fields. Therefore, it can be concluded that strong magnetic fields can 
affect the property of a non-extensive system by changing the magnetization significantly.

\begin{figure}[H]
\centering
\includegraphics[width=0.50\textwidth]{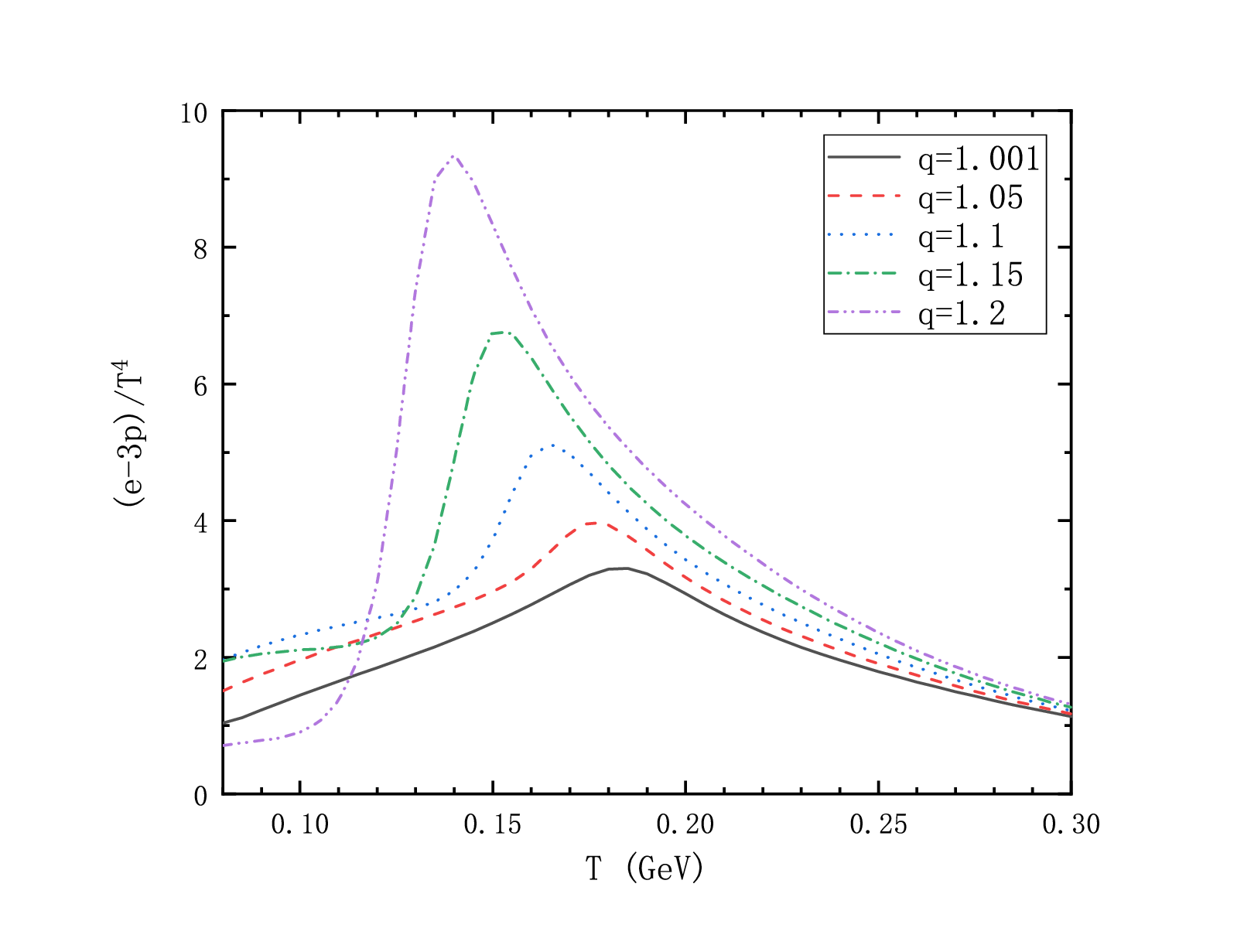}
\caption{The normalized interaction measure versus temperature is calculated for $q=1.001-1.2$, respectively.}\label{fig:Graph7}
\end{figure}

We all know that interaction measure is the trace of energy-momentum tensor and 
gives a deviation between the energy density and the pressure. Fig. \ref{fig:Graph7} 
illustrates how the interaction measure normalized by $T^4$ depends on the temperature 
for five sets of $q$-value at vanishing chemical potential. At low temperature, there is 
a visible rapid increase of the interaction measure because the energy density is rising 
faster than the pressure. A peak appears in the region around and just above 
the pseudo-critical temperature. The interaction measure vanishes at high temperature 
in the ideal gas limit. Compared with the $q=1.001$, the parameter $q=1.2$ results in 
more obvious trace anomaly, that is generated by the breaking of the conformal invariance 
of QCD at the quantum level.

\begin{figure}[H]
\centering
\includegraphics[width=0.50\textwidth]{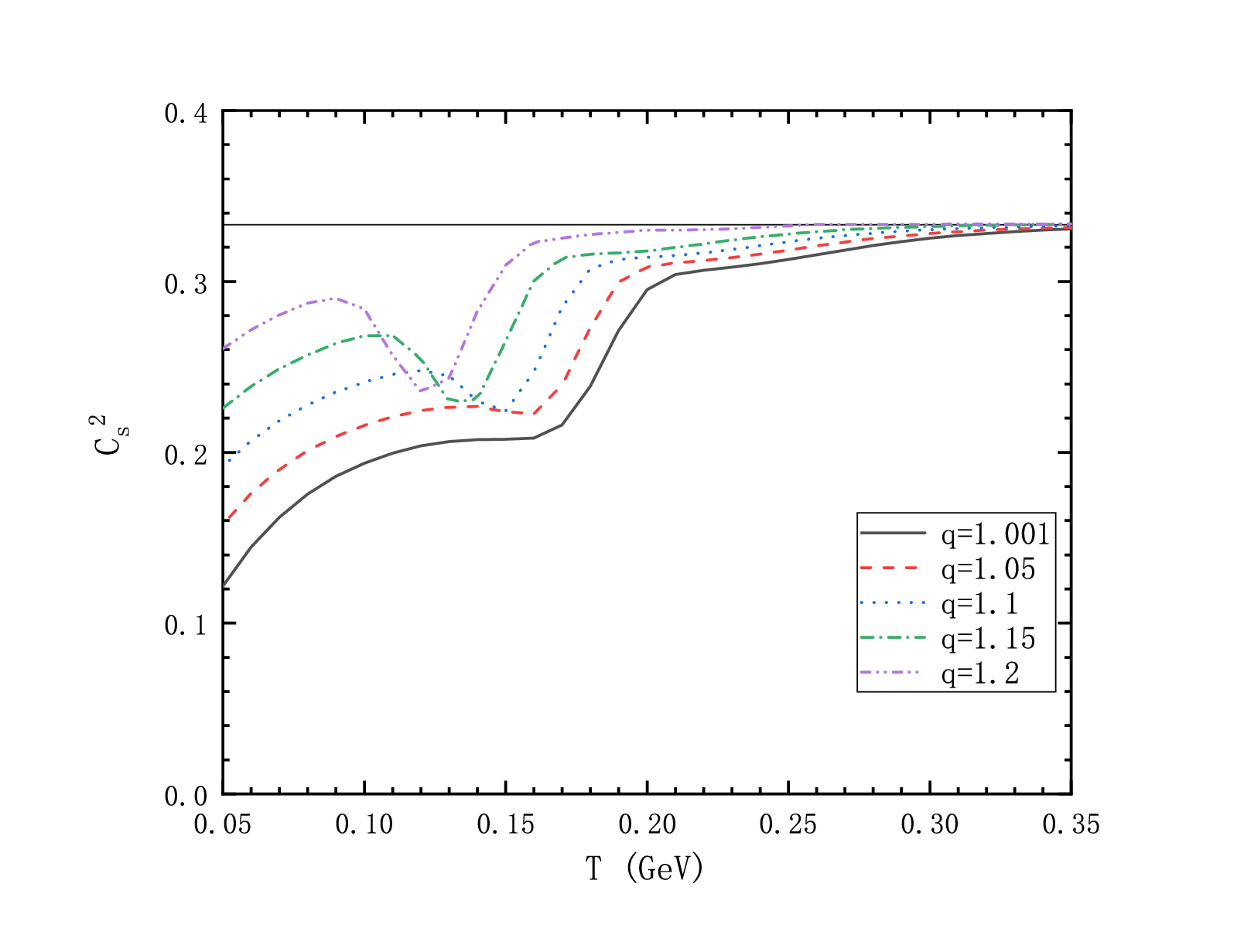}
\caption{The behavior of the speed of sound squared as a function of $T$ 
at the magnetic field $eB=0.2$ GeV$^2$.}\label{fig:Graph8}
\end{figure}

In Fig. \ref{fig:Graph8}, the speed of sound squared has been plotted as a function of 
the temperature for different $q$-parameter. The $c_{s}^{2}$ passes through a local minimum 
around the pseudo-critical temperature, and then grows up. As the $q$-value increases, 
the position of the dip moves to lower temperature. When the system is away from equilibrium, 
the speed of sound squared would have a large value. The curve displays a twist in the vicinity of 
the pseudo-critical temperature. At the temperature higher than $T_{pc}$, release of 
the degrees of freedom causes the speed of sound to increase rapidly and eventually reach 
the Stefan-Boltzmann limit at high temperature. It should be paid much attention that 
at higher temperatures more excited levels corrections might play a prominent role in 
determining the behavior of the speed of sound squared. When higher levels are included 
at large temperatures, it would asymptotically converge to the Stefan-Boltzmann limit value. 
The minimum value of $c_{s}^{2}$, also called the softest point or weak dip, is an important 
indicator of the transition in heavy ion collisions. It is obviously that the minimum moves 
towards lower temperature with the $q$ increases.

\section{Summary}
In this paper, we have investigated the three-flavor NJL model for hot QCD 
in the presence of strong magnetic fields to gain insight into the 
thermodynamical behavior of magnetized quark matter. The effect of non-equilibrium state 
has been considered by introducing a non-extensive parameter $q$. The definition of 
the distribution function in the NJL model has been reformulated using Tsallis statistic 
for fermions. The thermodynamic quantities have been obtained according to 
the q-exponential momentum distributions and the energy dispersion relations.

By changing the non-extensive parameter range from 1 to 1.2 in the presence of 
magnetic fields, we found that the system always undergoes a crossover transition 
for all given $q$ values, and the pseudo-critical temperature would decrease 
as the value of $q$ increases. Moreover, it is observed that the normalized pressure 
in the non-extensive system behaves as an increasing function of temperature 
as expected in literature. The system exhibits anisotropy due to the presence of 
strong magnetic fields, and the influence of the non-extensive parameter 
is more significant at high temperatures.

In particular, the magnetization plays an important role to understand quark matter, 
and its sign reflects whether QCD matter is paramagnetic or diamagnetic. It has been 
demonstrated that $\cal{M}$ is negative within a narrow temperature range 
for all $q$ values at zero chemical potential. The diamagnetic matter can be only realized 
at low temperature by increasing the magnitude of the magnetic fields. However, 
the paramagnetic property would be usually recovered at higher temperature. Finally, 
We showed that the behavior of thermodynamic quantities such as the interaction measure and 
the speed of sound squared changes more dramatically when the system is far from 
equilibrium state. In this process, we believed that the higher Landau levels also 
play a key role. As proposed in Ref. \cite{Avancini:2020xqe}, the $B$-dependent divergence is not 
subtracted (by renormalizing $B^2$) as in the MFIR. The vacuum magnetic regularization could 
avoid the unphysical result by an extra term to the pressure. The effect of the regularization shceme 
needs to be further studied in the future. We hope our conclusion is helpful for the investigation of 
non-equilibrium quark mater in heavy ion collisions experiment.

\acknowledgments{ The authors would like to thank the 
National Natural Science Foundation of China for 
support through Grants No.11875181, No.12047571, and 
No.11705163. This work was also sponsored by the Fund for 
Shanxi “1331 Project” Key Subjects Construction.}

\end{document}